# Reliability Quantification of Deep Reinforcement Learning-based Control

Hitoshi Yoshioka *      Hirotada Hashimoto **

Graduate School of Engineering, Osaka Metropolitan University

*su23152i@st.omu.ac.jp

**hashimoto.marine@omu.ac.jp

**Abstract**

Reliability quantification of deep reinforcement learning (DRL)-based control is a significant challenge for the practical application of artificial intelligence (AI) in safety-critical systems. This study proposes a method for quantifying the reliability of DRL-based control. First, an existing method, random noise distillation, was applied to the reliability evaluation to clarify the issues to be solved. Second, a novel method for reliability quantification was proposed to solve these issues. The reliability is quantified using two neural networks: reference and evaluator. They have the same structure with the same initial parameters. The outputs of the two networks were the same before training. During training, the evaluator network parameters were updated to maximize the difference between the reference and evaluator networks for trained data. Thus, the reliability of the DRL-based control for a state can be evaluated based on the difference in output between the two networks. The proposed method was applied to DQN-based control as an example of a simple task, and its effectiveness was demonstrated. Finally, the proposed method was applied to the problem of switching trained models depending on the state. Consequently, the performance of the DRL-based control was improved by switching the trained models according to their reliability.

**1. Introduction**

Deep reinforcement learning (DRL) has attracted considerable attention as a method for realizing autonomous control. It achieves high performance even for highly complicated tasks. However, the black-box problem prevents the practical application of DRL-based controls in safety-critical systems. In DRL, the agent, which is a subject of action, autonomously learns the action that maximizes the cumulative discounted reward using multi-layer neural networks. Therefore, the intentions and decision-making processes of DRL-based control remain unclear. It is difficult for human users to accept DRL-based control outputs without showing the reliability of the DRL-based control.

Furthermore, the causes of inappropriate DRL-based control decisions must be explained. Recently, the explainability of artificial intelligence (AI) has become crucial and has been studied as an explainable AI (XAI). One approach is to analyze the change in outputs by eliminating part of the input data or by adding noise. [1,2]. It can analyze the contributions of data/area in the input data and can be applied to AIs of any structure, such as random forest and neural networks. However, output changes due to many patterns of change for a specific input data/area should be analyzed to calculate the contribution. Therefore, the analysis could not be performed in real time because it required several hours. In another approach, the contribution of the input data is explained by an analysis using backpropagation. This method explains the contribution of the input data by differentiating the output of the AI with respect to the input data [3,4,5]. This approach can be applied to any model if it is differentiable. Furthermore, several methods suitable for image recognition have also been proposed [6,7,8,9,10]. These methods can visualize contributions of input data as a heatmap. Although they have excellent visibility, they can only be applied to convolutional neural networks (CNNs). Furthermore, AI learns the complex relationship between input and output data that humans cannot understand; therefore, visualized contributions are often unconvincing to humans. Consequently, many XAI methods have been proposed thus far [11,12,13,14]; nevertheless, they cannot sufficiently explain the AI's intentions, and evaluation methods for XAI themselves have not been proposed. An XAI method that can be applied to the AI of any structure in real-time and that provides convincing outputs for humans needs to be studied.

In addition to explainability, reliability is another important factor in XAI. During the operation of a system, it is necessary to confirm that the system is working properly. Automatic control based on model predictive control (MPC) can be evaluated based on whether the iteration of the optimization calculation is convergent. However, no method for evaluating the adequacy of AI-based controls has yet been proposed. In DRL, an agent learns a policy on how to act through experience; therefore, the reliability of DRL-based control relies exhaustively on whether the agent experiences the possible states in a learning environment. Because a dataset is sampled according to the agent's action and training data are randomly chosen from the sampled dataset, the sampled state has a bias caused by the agent's policy. Because it is difficult to learn all the states, there is a lack of learned states. Therefore, identifying the operational design domain (ODD) of AI requires numerous simulations covering all states in the learning environment. To solve this problem, a method that prioritizes the sampled data according to the learning loss was proposed. This can reduce training data bias when selecting the training data from the sampled dataset. However, this does not reduce the bias in the sampled dataset. To reduce this bias, improved exploration methods are necessary to enable the agents to experience various states. One approach to improving exploration is to provide an intrinsic reward to the agent when the agent experiences an unknown state. In random network distillation (RND) [15], two randomized neural networks, the target network and predictor network, are introduced to calculate the intrinsic reward. In the learning process, the parameters of the target network were fixed, and those of the predictor network were updated to minimize the error between the target and predictor networks. Because the error is minimized for well-experienced data, it can be used as an intrinsic reward. Therefore, curiosity can be implemented in a DRL agent through the RND. RND improves learning efficiency and is especially effective in cases with a sparse reward. Although several methods for reducing bias in training data have been studied, they cannot be eliminated. Therefore, quantitatively evaluating the bias of training data for the practical application of AI-based systems is crucial.

The simplest method for assessing whether the AI has learned the state well is to count the data used [16]. Because a state is defined in a continuous multidimensional space, the number of state patterns is infinite. Therefore, determining the number of counts necessary for learning in every situation is difficult. Moreover, the same idea as the RND can be used to evaluate proficiency. As RND induces the exploration of unknown states through intrinsic reward, the value of the intrinsic reward can describe the proficiency of the state. However, the applicability of the RND to proficiency evaluation is unclear.

In this study, a novel method for reliability quantification of DRL-based controls is proposed. Reliability is evaluated by the fact that learning has been performed sufficiently for the given states. First, RND was applied to the reliability evaluation to clarify the issues to be solved. Second, a reliability quantification method is proposed to solve these issues. The reliability is quantified using reference and evaluator networks, which have the same structure and initial parameters. During the training, the parameters of the evaluator networks were updated to maximize the difference between the reference and evaluator networks. Thus, the reliability of the DRL-based control for states can be evaluated based on the difference in outputs between the two networks. For example, it was applied to deep q-network (DQN)-based control for a simple task, and its effectiveness was demonstrated. Finally, the switching of the trained DQN models is demonstrated as an example of the application of the proposed reliability quantification.

## 2. Deep Reinforcement Learning

2.1. Reinforcement Learning

Reinforcement learning (RL) consists of an agent and environment. In the learning process, the agent observes the current state of the environment, $s_t$. The agent then takes action according to its policy, π, from an observed state. The environment transits to the next state, $s_{t+1}$, and returns a reward, $r_{t+1}$, to the agent. Finally, the agent updates its policy to maximize the cumulative reward in the future using the received reward. The cumulative reward $R_t$, is calculated using Eq. 1, where γ is a discount rate and shows an uncertain amount of reward obtained in the future.

$$R_t = r_{t+1} + \gamma r_{t+1} + \gamma^2 r_{t+2} \dots \tag{1}$$

Owing to the huge trial-and-error time, the agent obtains the optimal action that can maximize the cumulative reward.

2.2. Deep Q-Network

In this study, the deep Q-network (DQN) [17] was used to develop autonomous control. DQN is one of the DRL methods based on the Q function, $Q_\pi(s_t, a|\theta)$. The Q function maps the state and action to the value of the action. The action value is called the Q value, and the optimal action is that has the maximum Q value. In DQN, the neural network approximates the Q function. The parameters of the neural network were updated to minimize the loss function in Eq. 2, which indicates the error between the predicted Q value and the obtained value. The $Q_\pi^*(s, a|\theta')$, called target network, is a neural network that has the same structure as $Q_\pi(s, a|\theta)$. It is used when predicting future rewards and stabilizing training. The parameters of the target network, $\theta'$, are updated by Eq. 3, where $\alpha$ is a learning rate.

$$\text{Loss}_{\text{DQN}}(\theta) = \mathbb{E}\left[\left(r_{t+1} + \gamma \max_{a' \in A} Q_\pi^*(s_{t+1}, a'|\theta') - Q_\pi(s_t, a_t|\theta)\right)^2\right] \tag{2}$$

$$\theta' \leftarrow \theta' + \alpha(\theta - \theta') \tag{3}$$

**3. Learning Environment**

In this section, the learning environment for the validation of the existing and proposed methods is described.

3.1. Learning Task

The task of the agent was to achieve the goal from its initial position. The goal $(x_{\text{goal}}, y_{\text{goal}})$, is set at $(0,0)$ and the initial position of the agent $(x_{\text{init}}, y_{\text{init}})$, is randomly set in an area described by Eq. 4. The initial velocities $(v_{x_{\text{init}}}, v_{y_{\text{init}}})$, are $(0,0)$. The area of the learning environment is given by Eq. 5.

$$(x_{\text{init}}, y_{\text{init}}) \in \left\{(x, y) \middle| \left(200 \leq \sqrt{(x_{\text{goal}} - x)^2 + (y_{\text{goal}} - y)^2} \leq 300\right)\right\} \tag{4}$$

$$\{(x, y) | -400 \leq x \leq 400, -400 \leq y \leq 400\} \tag{5}$$

The motion of the agent is defined as the motion of a mass point, considering the resistance corresponding to the velocity. The equations of motion are described by Eqs. 6, where $m$, $f$, $\kappa$ are a mass, a force for each axis, and a gain of resistance, respectively. The values of mass and resistance gain were set to 10 and 2, respectively.

$$\begin{cases} m\dfrac{d^2x}{dt^2} = f_x - \kappa\dfrac{dx}{dt} \\ m\dfrac{d^2y}{dt^2} = f_y - \kappa\dfrac{dy}{dt} \end{cases} \quad (6)$$

As the action space of the DQN is defined in a discrete space, the actions of the AI are set as discrete forces, F. The action options are the nine forces described in Eqs. 7.

$$F = \begin{pmatrix} f_x \\ f_y \end{pmatrix} := \left\{ \begin{pmatrix} f\sin(\theta) \\ f\cos(\theta) \end{pmatrix} \middle| (f=0) \vee \left( f = 10 \wedge \theta \in \left\{ 0, \dfrac{\pi}{4}, \dfrac{\pi}{2}, \ldots, \dfrac{7\pi}{4} \right\} \right) \right\} \quad (7)$$

The schematic view of the learning environment is shown in Fig. 1. The number of steps in one episode was 240, and the time step was set to 1. The agent makes decisions at each time step. The conditions for ending the episode are as follows: 240 steps are performed, the distance between the agent and goal becomes less than 10, or the agent exits the learning environment.

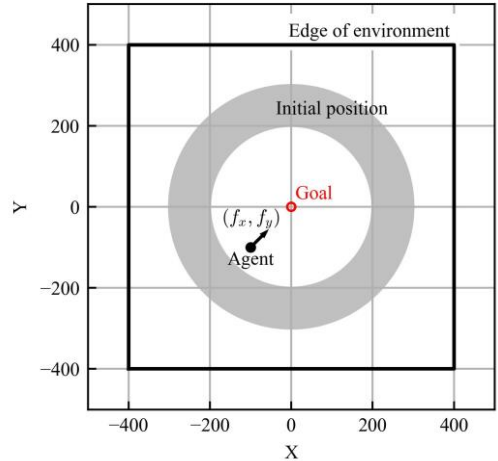

**Figure 1.** Learning environment

### 3.2. Design of Reward

Because the agent determines the goodness of an action using a reward, a reward appropriate to the task should be designed. To realize an action to achieve a goal, a reward is given according to the deviation from the goal, $r_t^{dev}$. Furthermore, the reward, according to the changes in the distance between the agent and the goal, $r_t^{dec}$, is given if the distance decreases. Finally, a large negative reward was provided if the agent exited the environment. The rewards and total reward were calculated using Eqs. 8, 9, 10, and 11, where $d_t$ is the distance between the agent and goal.

$$r_t^{dev} = -\dfrac{d_t}{400\sqrt{2}} - 0.1 \quad (8)$$

$$r_t^{dec} = 0.03 + 0.07 * \left| 1 - \arccos\left(\frac{d_{t+1} - d_t}{\sqrt{v_x^2 + v_y^2}}\right) \Big/ (\pi/2) \right| \qquad (9)$$

$$r_t^{done} = \begin{cases} -10, & \text{if the agent go out} \\ 0, & \text{else} \end{cases} \qquad (10)$$

$$r_t = \begin{cases} r_t^{dev} + r_t^{done}, & d_{t+1} > d_t \\ r_t^{dev} + r_t^{dec} + r_t^{done}, & d_{t+1} \le d_t \end{cases} \qquad (11)$$

3.3. Design of Observation

The observation is the input data for the neural network and should include sufficient data to predict the reward. In this study, the observed state is defined by Eq. 12, where $(dx, dy)$ and $(v_x, v_y)$ are the relative positions of the goal and the velocities of the agent, respectively. These values are scaled from -1 to 1. The input data for the neural network were the current state and the states in the previous four steps.

$$s_t := [v_x/5, v_y/5, dx/800, dy/800] \qquad (12)$$

**4. Reliability Quantification**

4.1. Applicability of Random Network Distillation

In random network distillation (RND), an intrinsic reward is provided when an agent experiences an unknown state, and exploration of the unknown state is encouraged. Thus, the value of the intrinsic reward was used as the value of inadequacy to the state, and RND is used for reliability evaluation. A neural network consists of multiple layers, and features are extracted from each layer. Because the output of each layer may be changed significantly by smaller changes in the input data, it is not possible to extract similar features from similar states. Therefore, the uncertainties of both the input state and the extracted feature are evaluated. The structure of the neural network is defined according to the RND, as shown in Fig. 2. The neural network is comprised of three parts: a DQN network, an RND network that evaluates the uncertainty of the state, and an RND network that evaluates the uncertainty of the extracted features. They consist of fully connected layers (FC); the numbers of layers and nodes are shown in Fig. 2. The activation functions in the DQN are the ReLU function in the hidden layers and the linear function in the output layer. Activations in the hidden layers of the RND networks and output layers are a softsign function and linear function. Any DRL method can be applied by setting up a neural network instead of a DQN.

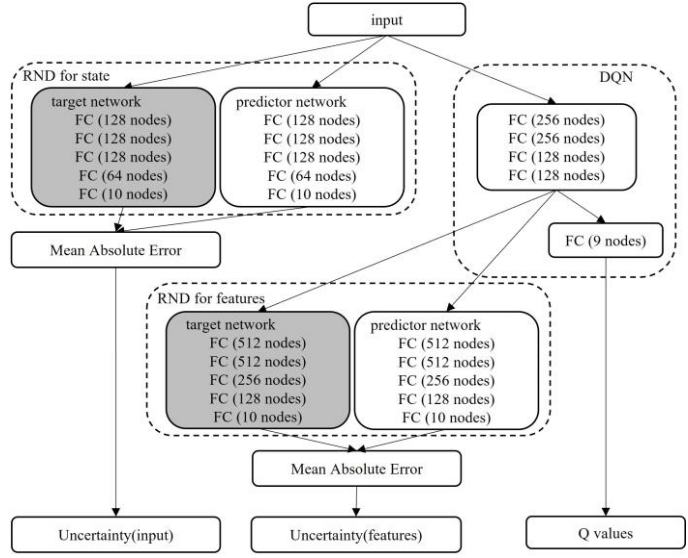

**Figure 2.** Structure of the neural networks considering RND

During training, the parameters of the neural networks that evaluate uncertainty are updated to minimize the output values. The parameters of the neural network predicting the Q values were updated to minimize the loss function of the DQN. To ensure that reliability quantification does not affect the DRLs, these updated processes are independent. The training hyperparameters are listed in Table 1.

**Table 1.** Hyperparameters for training

| | |
|---|---|
| Learning steps | 20,000,000 [steps] |
| Batch size | 1024 |
| Learning rate | 0.001 |
| Target model update | 0.01 |
| optimizer | Adam |
| L1 regularization | 0.0001 |

After training until the loss values converged, the trained model was evaluated in the simulation by varying the initial position within the ranges shown in Eqs. 13. To evaluate the action of the agent and the uncertainty of the learning environment, the ranges were wider than those in the learning environment, and the episode did not end if the agent exited the learning environment.

$$(x_{evalu}, y_{evalu}) \in \left\{(x, y) \left| \begin{array}{l} x \in \{-700, -630, \ldots, 630, 700\} \\ y \in \{-700, -630, \ldots, 630, 700\} \end{array} \right.\right\} \quad (13)$$

The evaluation trajectories are shown in Fig. 3. The unfilled and filled red markers indicate the agent's initial and last positions, respectively, and the shade of color indicates the time history. As shown in Fig. 3, the agent can reach the goal regardless of where the initial position is set within the learning environment. If the initial position is set outside the environment, the agent cannot reach the goal.

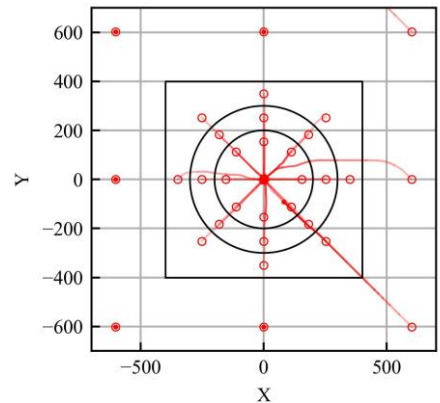

**Figure 3.** Trajectories of the trained model (RND)

Fig. 4 shows the distribution of the uncertainty. The uncertainty evaluated for input state and extracted features are denoted by 'input' and 'extracted features', respectively. The average of them is done by 'both'. The figure shows the average uncertainty values when the agent passed through the points. During the training process, the agent acted randomly in the early steps and experienced a wide range of states in the environment. At the end stage of training, the agent moves to the given goal from its initial point on the shortest course. Thus, the number of states experienced inside the initial position, inside, and outside the learning environments decreased in that order. According to Fig. 4, the value of uncertainty regarding the extracted features decreases with the distance to the goal. This result is appropriate for the number of experiences. However, the reliability of the input data is not suitable. Thus, the uncertainty can be evaluated using the RND with the extracted features. To compare the uncertainty distribution of the trained model with that of the untrained model, the uncertainty distribution of the untrained model is shown in Fig. 5. The uncertainty values were lower in some areas of the model distribution before training.

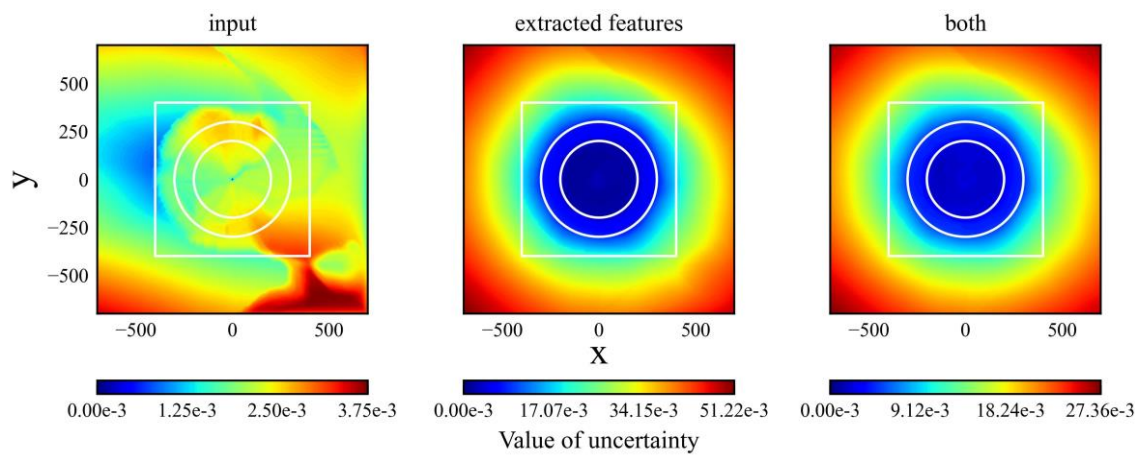

**Figure 4.** Distributions of uncertainty of trained model

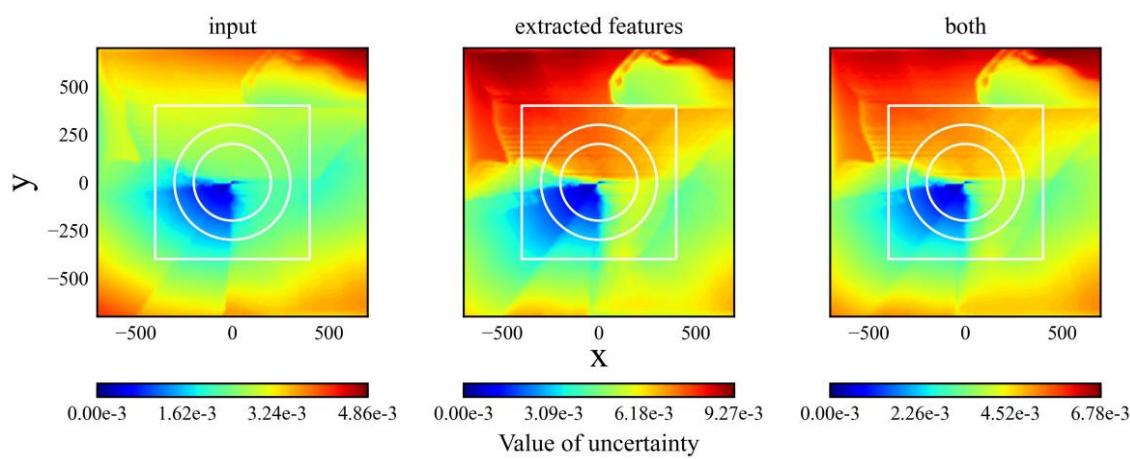

**Figure 5.** Distributions of uncertainty of model before training

Based on the results shown in Figs. 4 and 5, the possibility of applying RND to uncertainty quantification is shown; however, there are two issues regarding the application of RND to evaluate uncertainty. First, the RND does not ensure that the value of uncertainty is large in an unknown state, as shown in Fig. 5. Second, the uncertainty range depends on the initial parameters. If the maximum and minimum values differ according to the model, it is difficult to set a criterion to determine whether the model is in a well-trained state. These issues were caused by the value difference in the output of the target network, and the predictor network depended on the initial parameters. From the perspective of evaluation, it is important to ensure that the value of uncertainty is small in well-known states and large in unknown states and that the range of uncertainty is constant. If this is not ensured, the uncertainty coincidentally decreases for unknown states. Therefore, it is not preferable to use RND as an evaluation method.

4.2. Reliability Quantification Method

In the previous section, issues related to the application of RND to uncertainty evaluation were addressed. Thus, an improved method suitable for evaluating reliability was proposed. To evaluate reliability, two networks, a reference network, and an evaluator network, were used. Reliability was obtained by calculating the absolute error between the reference and evaluator networks, and a softsign function was applied to the absolute error to scale a range from 0 to 1. The reliability value was calculated using Eq. 14, and the calculation flow of reliability is shown in Fig. 6.

$$\text{Reliability}(s) = \text{softsign}(|\text{Ref}(s|\theta_r) - \text{Evalu}(s|\theta_e)|) \tag{14}$$

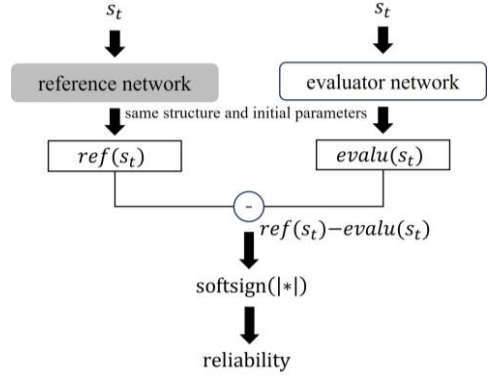

**Figure 6.** Reliability calculation flow

The scheme is similar to that of RND, but there are some differences. In RND, the target and predictor networks can have different structures, and their initial parameters are determined randomly; however, the reference and evaluator networks should have the same structure and initial parameters. Because the reference and evaluator networks are the same neural networks before training, their outputs are the same if the same input date is provided. This ensured that the reliability value reached zero before training. During the training process, the parameters of the reference network were fixed, and those of the evaluator network were updated to maximize the reliability value. Thus, the loss function can be described using Eq. 15, where $\text{Ref}(s|\theta_r)$, $\text{Evalu}(s|\theta_e)$, $\theta_r$, and $\theta_e$ are the reference network, the evaluator network, the parameters of the reference network, and the parameters of the evaluator network, respectively. The shape of the loss function is shown in Fig. 7.

$$\text{Loss}_{\text{improve}}(\theta_e) = \mathbb{E}[-\text{softsign}(|\text{Ref}(s|\theta_r) - \text{Evalu}(s|\theta_e)|)] \tag{15}$$

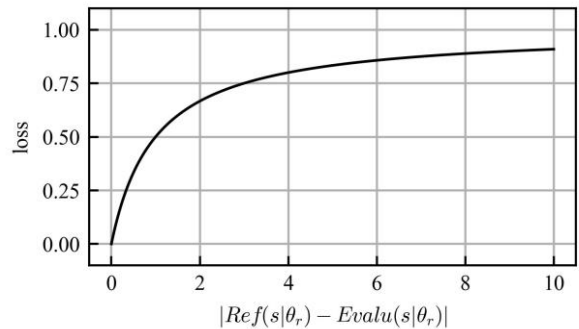

**Figure 7.** Shape of the loss function

The gradient of the loss function decreased as reliability increased. Therefore, the larger the reliability value, the more difficult it is to change. This means that sufficient trained experience fades slowly and is similar to human memory.

Furthermore, because the trained model is largely affected by recent training data, the reliability of past trained situations should decrease step-by-step. In this regard, the parameters of the evaluation network were updated to minimize the reliability of irrelevant training data, $S_{irr}$ and irrelevant data were generated randomly. The loss function that considers the forgetting experience is described by Eq. 16, where $s_{irr}$ is a randomly generated irrelevant states.

$$\text{Loss}_{forget}(\theta_e) = \mathbb{E}[\text{softsign}(|\text{Ref}(s_{irr}|\theta_r) - \text{Evalu}(s_{irr}|\theta_e)|)] \tag{16}$$

Finally, the loss function on reliability quantification is described as Eq. 17.

$$\text{Loss}_{reliability}(\theta_e) = \text{Loss}_{improve}(\theta_e) + \text{Loss}_{forget}(\theta_e) \tag{17}$$

The regularization term used to prevent overtraining is defined in Eq. 18, where p and λ are a dimension and a power of regularization, respectively. This restricted the distance of the parameters between the evaluation and reference networks.

$$\lambda \frac{1}{p} |\theta_e - \theta_r|^p \tag{18}$$

4.3. Validation of Reliability Quantification

To validate the proposed method, it was evaluated in the same manner as that described in Section 4.1. The structure of the neural network is shown in Fig. 8. The hyperparameters for training are the same as those listed in Table 1.

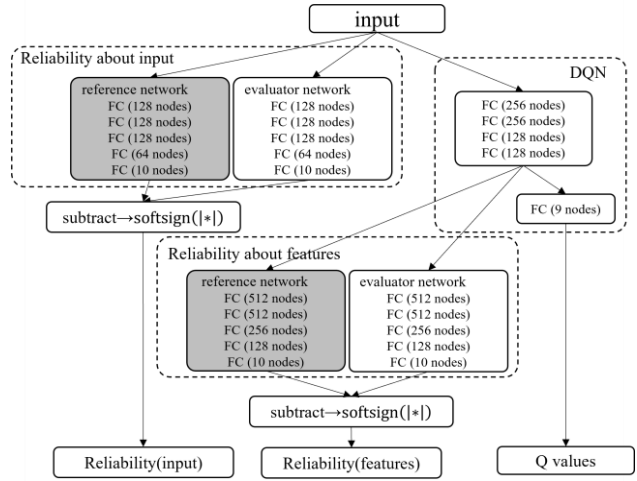

**Figure 8.** Structure of neural network on reliability quantification

The simulation trajectories are shown in Fig. 9. As shown in Fig. 9, the agent learns the optimal action within the learning environment. The reliability distribution is shown in Fig. 10. According to Fig. 10, the reliability becomes high within the range of the initial positions. According to the results, the proposed method, particularly the reliability of the extracted features, can evaluate whether the agent is trained.

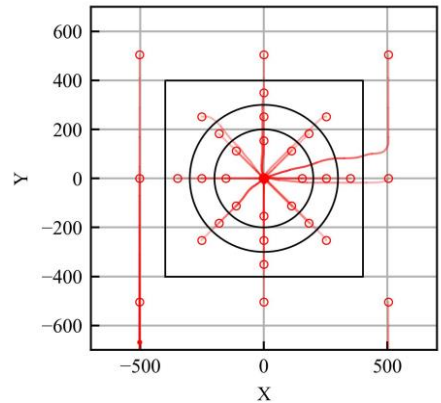

**Figure 9.** Trajectories of trained model (reliability quantification)

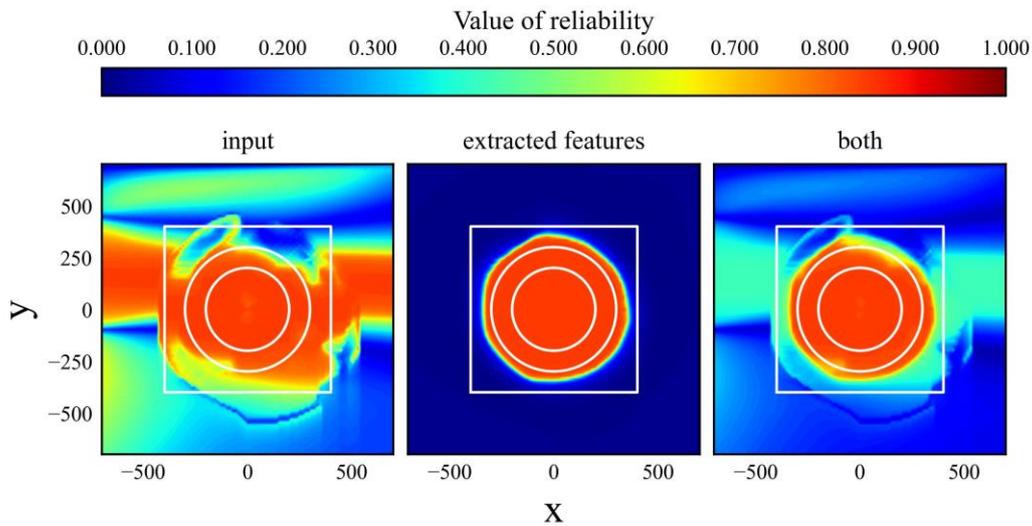

**Figure 10.** Distribution of reliability (trained model)

## 5. Discussion

5.1 Comparison with Random Noise Distillation

The proposed method was compared with uncertainty quantification using the RND. The distributions of the trained and untrained models are presented in Figs. 11 and 12, respectively. Because reliability and uncertainty are opposite evaluation indexes, the value size has the opposite meaning.

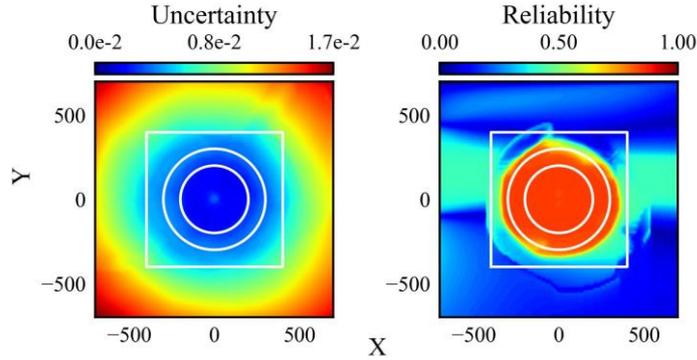

**Figure 11.** Comparison of distribution between uncertainty by RND and reliability by proposed (well-trained)

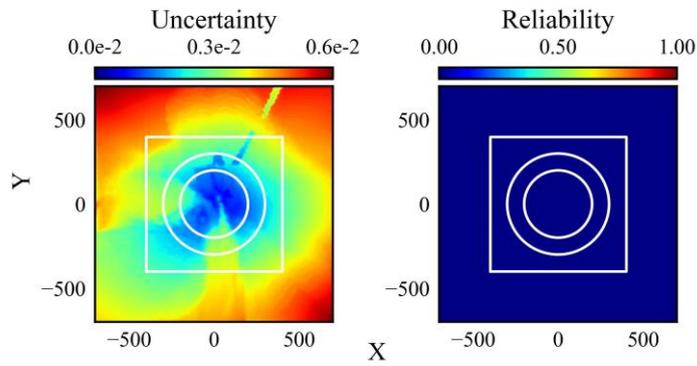

**Figure 12.** Comparison of distribution between uncertainty by RND and reliability by proposed (untrained)

According to the comparison results, the distribution clearly classifies the high-reliability and low-reliability areas rather than the uncertainty of RND. The range of the uncertainty distribution depends on the model parameters and structure in RND, whereas the range of the reliability distribution is fixed at 0–1 in the proposed method. Furthermore, as shown in Fig. 10, the reliability of the untrained model was 0 for all the states. These two aspects are important for their use as evaluation methods.

5.2 Relationships between Reliability of AI and Feasibility on Task

The agent must achieve the goal if the initial position is within a high-reliability area. The trajectories of the agent corresponding to the reliability at the initial position are shown in Fig. 13. The ratio for achieving this goal is shown in Fig. 14.

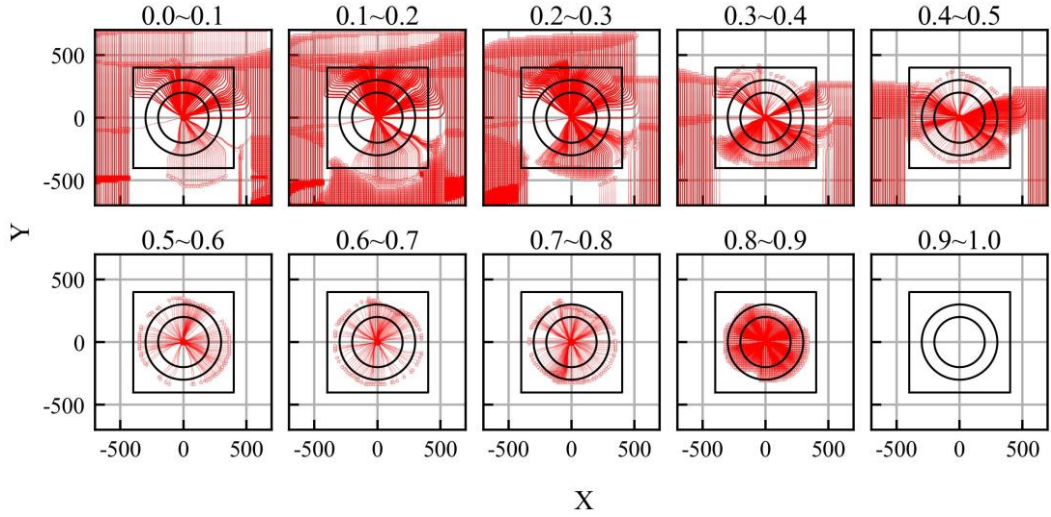

**Figure 13.** Trajectories corresponding to the reliability at the initial position

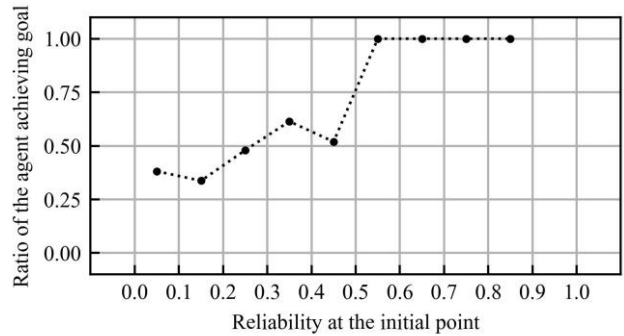

**Figure 14.** Ratio of the agent achieving its goal

According to the results, the ratio of the agent achieving its goal increases with the reliability value. If the reliability exceeds 0.5, the agent reaches the goal regardless of its initial position. Because the reliability and ratio of tasks achieved are shown, the proposed method can be used to quantify the reliability of DRL-based control.

**6. Switching of Trained Model**

It was demonstrated that the proposed method could evaluate the reliability of the trained model described in the previous section. The reliability value becomes 0 in the untrained state and 1 in the well-trained state. In RND, because the range of uncertainty depends on the model, a comparison of the uncertainty between models is not possible. The proposed method ensured that the reliability ranges were identical, enabling the reliability of the models to be compared. Therefore, the effectiveness of switching trained models was demonstrated as an application method. The quality of the sampled training data is a major issue in DRL. However, it is difficult to solve this problem completely. One solution is to switch between the trained models. Using several trained models, the untrained states of each model can cover each other. However, it is difficult to set criteria for switching models. As a solution, the reliability value can be used as a criterion for the switching model. Therefore, the effectiveness of switching models is discussed in this section.

Before demonstrating the switching models, four models with biased experience were developed. The ranges of the initial positions were different for these models. The initial positions of the i-th model were determined randomly within the range expressed in Eq. 19. The i-th model trained the states of the i-th quadrant. The reward and observations were the same as those described in the previous sections.

$$(x_{init}^i, y_{init}^i) \in \left\{ (d \cdot \cos(\theta), d \cdot \sin(\theta)) \middle| \begin{array}{c} 200 \leq d \leq 300 \\ \frac{\pi}{2}(i-1) \leq \theta < \frac{\pi}{2}i \end{array} \right\} \quad (i = 1,2,3,4) \quad (19)$$

After training, all models were evaluated. The trajectories and distribution of the reliability are shown in Figs. 15 and 16, respectively. The gray area in Fig. 15 indicates the range of the initial states. According to the results, the trained models achieved their goals and exhibited high reliability within their trained quadrants.

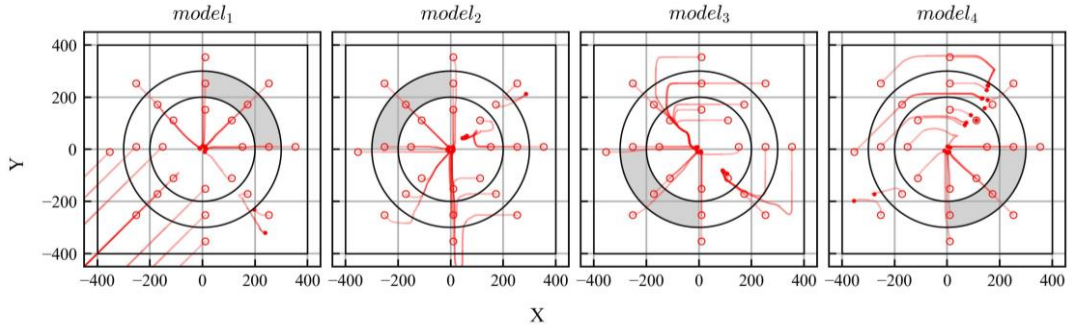

**Figure 15.** Trajectories of models trained using biased experience

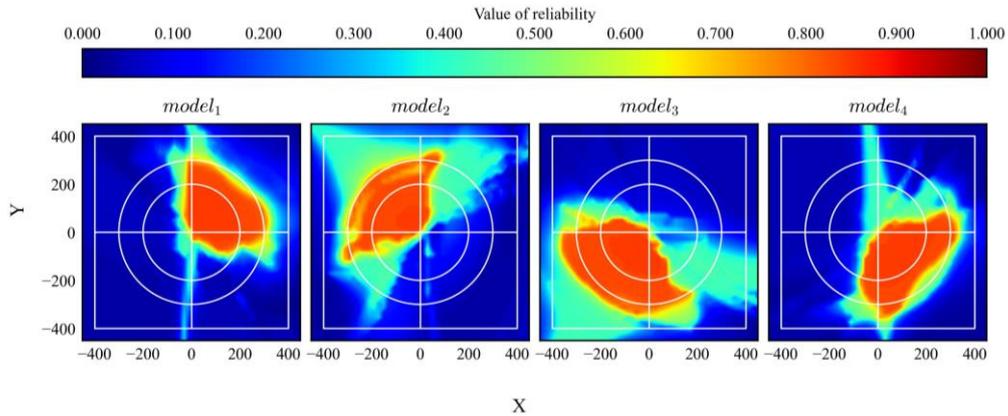

**Figure 16.** Reliability distributions of models trained using biased experience

The trained models were then switched. In the simulation, the optimal actions and reliability values of all models were calculated every timestep. Then, the action of the agent with the maximum reliability is used as an action in the simulation. The results of the trajectories and reliability are shown in Fig. 17. According to the evaluation results, the agent achieved the goal from all initial positions and exhibited high reliability over a wider area in the learning environment. Thus, the performance of the DRL-based control is improved by switching the trained models according to reliability.

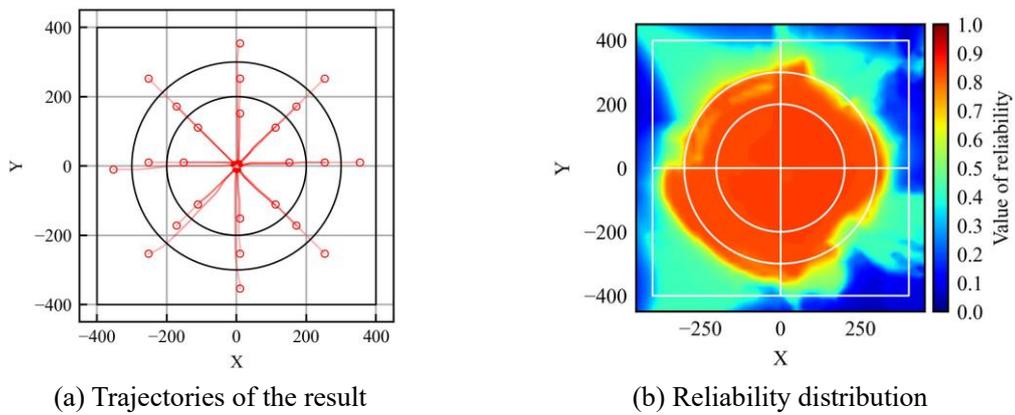

(a) Trajectories of the result  (b) Reliability distribution

**Figure 17.** Trajectories and reliability of results of model switching

## 7. Conclusion

A major challenge of DRL is reliability quantification. It is necessary to evaluate the reliability in real time for the actual application of DRL-based control in safety-critical systems. In this study, the following was conducted to propose a novel method for reliability quantification of DRL-based control. First, the applicability of an existing method, RND, for uncertainty evaluation is investigated. Uncertainty is defined as the opposite evaluation index of reliability. It was confirmed that the range of the uncertainty value depends on the initial parameters. Hence, it cannot be ensured that the uncertainty value increases in an unknown state. Because the range of uncertainty is not fixed and the uncertainty value often becomes small for untrained states, RND is difficult to use for uncertainty quantification. Second, a reliability quantification method is proposed to solve those problems. Reliability is evaluated by the difference between the reference and evaluator networks, which have the same structure and initial parameters. The parameters of the reference network were fixed, whereas those of the evaluator network were updated to maximize the difference in output between the two networks for the trained data and minimize the difference in output between them for the irrelevant data. The proposed method was validated for DQN-based control of a simple task. Consequently, it was confirmed that the proposed method can evaluate reliability and identify a well-trained domain in the learning environment. Finally, an example application is presented. To address the lack of experience in a trained model, switching the trained models according to their reliability was investigated. Using four trained models with biased experience, it was demonstrated that a given task could be completed in any situation by appropriately switching between them based on their reliability. The advantages of the proposed method are that the range of values is fixed and that the value of reliability becomes zero in untrained situations and one in well-trained situations. These advantages are beneficial for evaluating reliability and creating a criterion easily. The proposed method therefore can be used in various applications, such as the switching of trained models. The issue of bias in experienced states can be resolved by switching several trained models. This can improve the performance and robustness of DRL-based control. Another application example involves identifying the ODD of the control. Additionally, the proposed method can calculate the intrinsic reward instead of the RND.

In future studies, its applicability to an actual environment should be validated. In actual environments, the input state is affected by data noise. Their effects are not clear at this moment but are important for practical use. The proposed method assumes that the loss value converges, although it does not ensure that the loss value in all states converges to zero. Therefore, the convergence of the loss of the training process in reliability quantification should be considered.